\documentstyle[11pt,aaspp4,flushrt]{article}
\def\etal{{et al.}}

\begin{document}

\title{The Discovery of a Second Field Methane Brown Dwarf from Sloan
Digital Sky Survey Commissioning Data\footnote{Based on observations
obtained with the Sloan Digital Sky Survey and the Apache Point
Observatory 3.5-meter telescope, which are owned and operated by the
Astrophysical Research Consortium, and with the United Kingdom
Infrared Telescope.}}  

\author{ 
     Zlatan I.\ Tsvetanov\altaffilmark{\ref{JHU}},
     David A.\ Golimowski\altaffilmark{\ref{JHU}}, 
     Wei Zheng\altaffilmark{\ref{JHU}},
     T.\ R.\ Geballe\altaffilmark{\ref{Gemini}}, 
     S.\ K.\ Leggett\altaffilmark{\ref{UKIRT}},
     Holland C.\ Ford\altaffilmark{\ref{JHU}},
     Arthur F.\ Davidsen\altaffilmark{\ref{JHU}},
     Alan Uomoto\altaffilmark{\ref{JHU}},
     Xiaohui Fan\altaffilmark{\ref{Princeton}}, 
     G.\ R.\ Knapp\altaffilmark{\ref{Princeton}}, 
     Robert H. Lupton\altaffilmark{\ref{Princeton}},
     Jeffrey R. Pier\altaffilmark{\ref{Flagstaff}}, 
     Michael A. Strauss\altaffilmark{\ref{Princeton}}, 
     James Annis\altaffilmark{\ref{Fermilab}},
     J.\ Brinkmann\altaffilmark{\ref{APO}},
     Istv\'an Csabai\altaffilmark{\ref{JHU},\ref{Eotvos}}, 
     Robert B. Hindsley\altaffilmark{\ref{USNO}}, 
     \v{Z}eljko Ivezi\'{c}\altaffilmark{\ref{Princeton}}, 
     D.Q.\ Lamb\altaffilmark{\ref{Chicago}}, 
     Jeffrey A.\ Munn\altaffilmark{\ref{Flagstaff}}, 
     Heidi Jo Newberg\altaffilmark{\ref{Fermilab}}, 
     Ron Rechenmacher\altaffilmark{\ref{Fermilab}}, 
     Donald P.\ Schneider\altaffilmark{\ref{PennState}}, 
     Aniruddha R.\ Thakar\altaffilmark{\ref{JHU}}, 
     Patrick Waddell\altaffilmark{\ref{Washington}}, 
     and Donald G.\ York\altaffilmark{\ref{Chicago}}
     (the SDSS Collaboration) }

\newcounter{address}
\setcounter{address}{2}
\altaffiltext{\theaddress}{Department of Physics and Astronomy, 
     The Johns Hopkins University, 3701 San Martin Drive, 
     Baltimore, MD 21218, USA\label{JHU}}
\addtocounter{address}{1}
\altaffiltext{\theaddress}{UKIRT, Joint Astronomy Centre, 660 North
	A'ohoku Place, Hilo, HI 96720\label{UKIRT}}
\addtocounter{address}{1}
\altaffiltext{\theaddress}{Gemini North Observatory, 670 North
	A'ohoku Place, Hilo, HI 96720\label{Gemini}}
\addtocounter{address}{1} 
\altaffiltext{\theaddress}{Princeton University Observatory, 
	Princeton, NJ 08544\label{Princeton}}
\addtocounter{address}{1}
\altaffiltext{\theaddress}{U.S. Naval Observatory, Flagstaff Station, 
	P.O. Box 1149, Flagstaff, AZ  86002-1149\label{Flagstaff}}
\addtocounter{address}{1}
\altaffiltext{\theaddress}{Fermi National Accelerator Laboratory, 
	P.O. Box 500, Batavia, IL 60510\label{Fermilab}}
\addtocounter{address}{1}
\altaffiltext{\theaddress}{Apache Point Observatory, P.O. Box 59,
	Sunspot, NM 88349-0059\label{APO}}
\addtocounter{address}{1}
\altaffiltext{\theaddress}{Department of Physics of Complex Systems,
	E\"otv\"os University, P\'azm\'any P\'eter s\'et\'any 1/A, 
	Budapest, H-1117, Hungary\label{Eotvos}}
\addtocounter{address}{1}
\altaffiltext{\theaddress}{U.S. Naval Observatory, 3450 Massachusetts Ave., 
	NW, Washington, DC  20392-5420\label{USNO}}
\addtocounter{address}{1}
\altaffiltext{\theaddress}{University of Chicago, Astronomy \& Astrophysics 
	Center, 5640 S. Ellis Ave., Chicago, IL 60637 \label{Chicago}}
\addtocounter{address}{1}
\altaffiltext{\theaddress}{Department of Astronomy and Astrophysics, The 
	Pennsylvania State University, University Park, PA 16802\label{PennState}}
\addtocounter{address}{1}
\altaffiltext{\theaddress}{University of Washington, Department of Astronomy,
     Box 351580, Seattle, WA 98195\label{Washington}}
\addtocounter{address}{1}
\altaffiltext{\theaddress}{Ohio State University, Department of Astronomy,
	140 W.\ 18th Ave., Columbus, OH 43210\label{Ohio}}

\begin{abstract}
We report the discovery of a second field methane brown dwarf from the
commissioning data of the Sloan Digital Sky Survey (SDSS). The object,
SDSS~J134646.45--003150.4 (SDSS~1346--00), was selected
because of its very red color and stellar appearance.  Its spectrum
between 0.8--2.5~$\mu$m is dominated by strong absorption bands of
H$_2$O and CH$_4$ and closely mimics those of Gliese~229B and
SDSS~162414.37+002915.6 (SDSS~1624+00), two other known methane brown
dwarfs.  SDSS~1346--00 is approximately 1.5 mag fainter than
Gliese~229B, suggesting that it lies about 11~pc from the sun.  The 
ratio of flux at 2.1~\micron\ to that at 1.27~\micron\ is larger for 
SDSS~1346--00 than for Gliese~229B and SDSS~1624+00, which suggests 
that SDSS~1346-00 has a slightly higher effective temperature than the 
others.  Based on a search area of 130~deg$^2$ and a detection limit 
of $z^* = 19.8$, we estimate a space density of 0.05~pc$^{-3}$ for 
methane brown dwarfs with $T_{\rm eff} \sim 1000$~K in the 40~pc$^3$ 
volume of our search.  This estimate is based on small-sample 
statistics and should be treated with appropriate caution.
\end{abstract}


\section{INTRODUCTION}

Over the last five years, the study of brown dwarfs has evolved from a
theoretical notion to a long-awaited discovery to the classification
and modeling of a rapidly increasing known population.  Dozens of
candidate and {\it bona fide} brown dwarfs have now been identified
using a variety of techniques, including coronagraphic imaging of
nearby stars (Nakajima et al.\ 1994), spectroscopic tests for
primordial lithium (Basri, Marcy, \& Graham 1996; Rebolo et al.\
1996), searches of young open clusters (Hambly 1998, and references
therein), optical and near-infrared sky surveys (Tinney et al.\ 1998;
Kirkpatrick et al.\ 1999; Strauss et al.\ 1999; Burgasser et al.\
1999), and deep-field studies (Cuby et al.\ 1999).  Until recently,
the coolest known brown dwarf was Gliese~229B, a companion to a nearby
M1 dwarf (Nakajima et al.\ 1995).  The spectrum of Gliese~229B was
singularly remarkable for its exhibition of $H$- and $K$-band
absorption features attributable to methane (Oppenheimer et al.\ 1998;
Geballe et al.\ 1996).  Under equilibrium conditions, methane (CH$_4$)
becomes the dominant carbon-bearing molecule for $T_{\rm eff} <
1200$~K (Fegley \& Lodders 1996; Burrows et al.\ 1997).  Models of
Gliese~229B's infrared spectrum indicate an effective temperature of
900--1000~K for the brown dwarf (Allard et al.\ 1996; Marley et al.\
1996; Tsuji et al.\ 1996; Leggett et al.\ 1999).

Strauss et al.\ (1999) recently reported the discovery of a
Gliese~229B-like brown dwarf from spectroscopic observations of a
candidate identified from commissioning data of the Sloan Digital Sky
Survey (SDSS).  The optical and near-infrared spectrum of this object,
SDSSp~J162414.37+002915.6 (hereafter SDSS~1624+00), exhibits strong
absorption by H$_2$O and CH$_4$ and closely mimics the spectrum of
Gliese~229B.  Unlike Gliese~229B, which is a companion to a nearby
star, SDSS~1624+00 is isolated in the field.  Assuming that
SDSS~1624+00 has an effective temperature and luminosity identical to
those of Gliese~229B, Strauss et al.\ estimated a distance of 10 pc to
SDSS~1624+00.

Within three weeks of the discovery of SDSS~1624+00, a second field
methane brown dwarf, SDSSp~J134646.45--003150.4 (hereafter
SDSS~1346--00), was discovered from the same SDSS commissioning
data. (SDSS uses J2000 coordinates, and ``p'' indicates that the
astrometric solution is preliminary.) Shortly thereafter, the
discoveries of five more methane brown dwarfs were announced -- four
from the Two-Micron All Sky Survey (2MASS; Burgasser et al.\ 1999) and
one from the New Technology Telescope (NTT) Deep Field project (Cuby
et al.\ 1999).  In this {\it Letter}, we report the discovery of
SDSS~1346--00, present the first medium-resolution $J$-band spectrum
of a methane brown dwarf, and comment on the space density of methane
brown dwarfs in the solar neighborhood.

\section{OBSERVATIONS}


The Sloan Digital Sky Survey project is described in detail by Gunn \&
Weinberg (1995\footnote{see {\tt http://www.astro.princeton.edu/PBOOK/}}). 
Here, we briefly outline the characteristics of SDSS that are relevant 
to this work.

SDSS images are obtained with a very large format CCD camera
(\cite{Gunn98}) attached to a dedicated 3$^{\circ}$ field of view
2.5~m telescope\footnote{see {\tt
http://www.astro.princeton.edu/PBOOK/telescop/telescop.htm}} at the
Apache Point Observatory, New Mexico. The sky is imaged in drift-scan
mode through five broa-dband filters spanning 0.33--1.05~$\mu$m: $u'$,
$g'$, $r'$, $i'$, and $z'$, with central wavelengths/effective widths
of 3540\AA/599\AA, 4770\AA/1379\AA, 6222\AA/1382\AA, 7632\AA/1535\AA\
and 9049\AA/1370\AA, respectively (\cite{Fukugita96}).  The exposure
time in each band is 54.1 s. The photometric calibration is obtained
through contemporaneous observations of a large set of standard stars
with an auxiliary $20''$ telescope\footnote{see {\tt
http://www.astro.princeton.edu/PBOOK/photcal/photcal.htm}} at the same
site.  The data is processed through an automated pipeline at the
Fermi National Accelerator Laboratory, where the software performs
photometric and astrometric calibrations, and finds and measures
properties of all objects in the images\footnote{see {\tt
http://www.astro.princeton.edu/PBOOK/datasys/datasys.htm}}.



To date, a number of 2\fdg5-wide strips centered on the Celestial
Equator have been imaged as part of the SDSS commissioning
program. The region spanning right ascensions $12^{h}$ and $16^h30^m$
has been imaged twice, first in June 1998 and again in March 1999.
Parts of the region not imaged in June~1998 were imaged twice in March
1999.  Because the network of primary standard stars was not fully
established during commissioning, the absolute photometric calibration
of these images remains uncertain at the 5\% level.  Cross-correlating
the data from the twice-imaged region removes any uncertainty
regarding the identification of faint and very red objects, especially
those detected in only one bandpass.  Using this technique, we
identified all point sources with $i^*-z^* > 2.5$, including sources
detected through $z'$ only.  (See \S2.2 for explanation of
superscripts.)  After inspecting the images of each source, we found
that the two reddest sources, SDSS~1346--00 and SDSS~1624+00, were
also the most credible brown dwarf candidates.  SDSS 1346--00 was
detected only in the $z'$ images recorded on UT 1999 March~20 and 22.
We note that the astrometric positions of SDSS~1346--00 in the two
runs are consistent with one another.  Figure~1 shows the finding
chart for SDSS~1346--00.


Table 1 lists the SDSS magnitudes and uncertainties for SDSS~1346--00.
We indicate the preliminary photometric measurements with asterisks,
but retain the primes for the filters themselves.  The SDSS magnitudes
are in the AB system (\cite{Fukugita96}) and are given as asinh values
(\cite{Lupton99}). The $u^*$, $g^*$, $r^*$, and $i^*$ values all 
represent non-detections -- 5$\sigma$ detections of a point source with 
1$''$ FWHM images correspond to $u^* = 22.3$, $g^* = 23.3$, $r^* = 23.1$, 
$i^* = 22.5$, and $z^* = 20.8$. The two $z^*$ measurements for 
SDSS~1346--00 agree to within 1 $\sigma$. Its $i^*$--$z^* \sim 4$ is
consistent with the $i^*$--$z^* = 3.77 \pm 0.21$ measured for
SDSS~1624+00 (Strauss et al.\ 1999).  Note that M and L dwarfs are not
expected to be redder than $i^*$--$z^* \sim 2.5$ (\cite{Fan2000}). 


Near-infrared photometry of SDSS~1346--00 was obtained on UT 1999
May~23 using the IRCAM $256\times 256$ InSb array and the United
Kingdom Infrared Telescope (UKIRT). The plate scale was 0\farcs28
pixel$^{-1}$, and the exposure times at $J$, $H$, and $K$ were 5 min,
14 min, and 18 min, respectively.  The conditions were photometric,
and the seeing was 0\farcs8.  The object was imaged using the standard
dither technique, and the images were calibrated using observations of
UKIRT faint standards (\cite{Casali92}).  The UKIRT magnitudes of
SDSS~1346--00 are listed in Table~1.  (Vega has magnitude zero in all
UKIRT bandpasses.)  The $J-K$ and $H-K$ colors of SDSS~1346--00 are
redder by $\sim 0.1$~mag than those of Gliese~229B (\cite{Leggett99})
and SDSS~1624+00 (Strauss et al.\ 1999).  The $z^*-J$ color is about 
the same for two SDSS methane dwarfs. SDSS~1346--00 is fainter than 
Gliese~229B and SDSS~1624+00 by $\Delta J = 1.5$ and
$\Delta J = 0.3$, respectively.


An optical spectrum of SDSS~1346--00 was obtained on UT 1999 May~10
using the Double Imaging Spectrograph (DIS) on the Apache Point 3.5~m
telescope.  The spectra were taken using the low resolution gratings,
providing a spectral coverage of 0.4--1.05~$\mu$m with dispersions of
6.2 \AA\ pixel$^{-1}$ on the blue side and 7.1 \AA\ pixel$^{-1}$ on
the red side, and a 2\farcs0 slit.  The exposure time was 30 minutes.
The conditions were non-photometric, and the seeing was $\sim
1$\farcs5.  The initial flux calibration and removal of atmospheric
absorption bands were achieved through observations of the
spectrophotometric standard BD$\,$+26$\,^\circ$2606 (F subdwarf,
\cite{OkeGunn83}) over several nights.  The final flux calibration,
however, was obtained by matching the optical spectrum with the
near-infrared spectrum in the overlapping region near 1~$\mu$m (see
below).

The calibrated optical spectrum is included in Fig.~2.  Although the
spectrum is significantly noisier than that of SDSS~1624+00 (Strauss
et al.\ 1999), it shows similar characteristics.  The spectrum rises
steeply toward the near-infrared, and its shape matches the SDSS
photometry well.  A distinct H$_2$O absorption band centered at $\sim
0.94~\mu$m remains after subtraction of the telluric absorption feature
at the same wavelength.  No flux was detected shortward of $\sim
0.8~\mu$m.

Spectra covering the $J$, $H$, and $K$ bands were obtained on the
nights of UT 1999 May~23 and UT 1999 June~2 with the facility grating
spectrometer CGS4 (\cite{Mountain90}) at UKIRT.  The instument was
configured with a 300 mm camera, a 40 l mm$^{-1}$ grating, and a
256$\times$256 InSb array.  The 1\farcs2 slit projected onto two
detector pixels, providing a spectral resolving power $R$ in the range
300 to 500.  The $JHK$ spectral range was spanned by five overlapping
spectra with the following central wavelengths and total exposure
times: 0.95~$\mu$m (48~min), 1.1~$\mu$m (56~min), 1.4~$\mu$m (33~min),
1.8~$\mu$m (48~min), and 2.2~$\mu$m (21~min).  The individual spectra
were obtained by nodding the object 7\farcs32 (12 detector rows) along
the slit.  The final co-added spectrum has a resolution of
0.0025~$\mu$m across the $J$ and $H$ bands, and 0.0050~$\mu$m in the
$K$ band.  Spectra of Kr, Ar, and Xe lamps were used for wavelength
calibration, and is accurate to $\leq 0.001~\mu$m. Spectra of bright F
dwarfs were obtained repeatedly throughout the observations for
initial flux calibration (after removal of prominent H absorption
features) and subtraction of telluric absorption lines.  The
individual spectra were then combined and scaled to match the
near-infrared photometry.  The resultant spectrum is shown in Fig.~2.

On UT 1999 June~7 and June~10, we obtained two higher resolution (150
l mm$^{-1}$ grating, ${\rm R} \sim 3000$) CGS4 spectra of
SDSS~1346--00 over the wavelength region $1.235 < \lambda <
1.290~\mu$m.  The observing technique was similar to the one described
above, with a total exposure time of 52 min.  This wavelength region
spans the peak of the emergent energy spectrum of the brown dwarf.
The inset in Fig.~2 shows a smoothed (by 1.5 pixel) average of the two
spectra. The resolution of the smoothed spectrum is $\sim
0.0005~\mu$m, which is the highest yet reported for a cool brown
dwarf.  The individual spectra, including the many narrow lines at the
red end of the spectrum, matched well before being combined.  The
error bars for the resultant spectum are $\sim$ 7\% everywhere, but
increase to about twice that value near $1.27~\mu$m, where telluric
lines of \ion{O}{1} are strong and variable.  The two broad absorption
features at $1.243~\mu$m and $1.252~\mu$m are due to \ion{K}{1}
(Kirkpatrick et al.\ 1993).  

\section{DISCUSSION}

The 0.8--2.5~\micron\ spectrum of SDSS~1346--00 looks astonishingly
like that of Gliese~229B, as recalibrated by Leggett et al.\ (1999),
and that of SDSS~1624+00 (Strauss et al.\ 1999).  Strong absorption
bands of H$_2$O and CH$_4$ dominate the spectrum, and the absorption
lines of H$_2$O at 2.0--2.1~$\mu$m discussed by Geballe et al.\ (1996)
are also apparent.  Note that while the zero-point of Gliese~229B's
spectrum is slightly uncertain due to a possible miscorrection for
scattered light from Gliese~229A, no such uncertainty exists for our
spectrum.  Flux is not detected at the bottom of the H$_2$O band at
1.36--1.40~$\mu$m, but is detected in the deepest parts of the H$_2$O
bands at 1.15~$\mu$m and 1.8--1.9$~\mu$m and the CH$_4$ band at
2.2--2.5~$\mu$m.

The only significant differences between the spectrum of SDSS~1346--00
and those of Gliese~229B and SDSS~1624+00 are SDSS~1346--00's
somewhat stronger absorption lines of \ion{K}{1} at 1.2436~$\mu$m and
1.2536~$\mu$m and the slight excess of flux around 1.7~$\mu$m and
2.1~\micron.  The latter excess is also reflected in the slightly
redder $J$--$K$ and $H$--$K$ colors of SDSS~1346--00 compared with
those of SDSS~1624+00 and Gliese~229B. Figure~3 illustrates the
differences between the $K$-band spectra of these three methane brown
dwarfs.  Burgasser et al.\ (1999) have also noted differences in the
$H$-to-$K$ flux ratios of the 2MASS ``T'' dwarfs and Gliese~229B, and
they use these ratios to establish a preliminary spectral sequence for
those five brown dwarfs.  Following their example, we infer that
SDSS~1346--00 is somewhat warmer than SDSS~1624+00 and Gliese~229B.
However, accurate modelling of the spectra is required to confirm and
calibrate this assessment.

The widths of the \ion{K}{1} absorption doublet (EW~$\approx$ 6 and
9~\AA, FWHM = $820 \pm 50$~km~s$^{-1}$) correspond to a rotation rate
that greatly exceeds the escape velocity from even the most massive
brown dwarf.  Thus, rotational broadening alone cannot account for the
width of the \ion{K}{1} lines.  As the dust-free photospheres of cool
brown dwarfs are transparent in this wavelength range to depths with
brightness temperatures of $\sim 1700$~K (Matthews et al.\ 1996) and
pressures of $\sim 30$ bar (Marley et al.\ 1996), the
observed widths of the \ion{K}{1} doublet are probably caused by
pressure broadening.  Accurate modelling of this higher-resolution
spectrum should significantly constrain the gravity, temperature, and
pressure profiles of cool brown dwarfs.


Although the S/N of the optical spectrum of SDSS~1346--00 is
insufficient for a rigorous assessment, the overall shape of the
continuum may be linked to the pressure-broadened \ion{K}{1} lines.
The optical spectrum of SDSS~1346-00 is remarkably similar to those of
Gliese~229B (Schultz et al.\ 1998; Oppenheimer et al.\ 1998) and SDSS
1624+00 (Strauss et al.\ 1999).  Gliese~229B's optical flux is lower
by 1--2 dex than the fluxes predicted by the models of dust-free
photospheres that reproduce well its near-IR spectrum
(Schultz et al.\ 1998; Golimowski et al.\ 1998).  Possible
explanations of this large discrepancy include absorption by aerosols
produced photochemically by radiation from Gliese~229A (Griffith,
Yelle, \& Marley 1998), a warm dust layer deep in the photosphere
(Tsuji, Ohnaka, \& Aoki 1999), and extreme pressure-broadening of the
\ion{K}{1} doublet at $0.76~\mu$m (Tsuji, Ohnaka, \& Aoki 1999;
Burrows, Marley, \& Sharp 1999).  The similarity between the
optical spectra of Gliese~229B and the SDSS field methane dwarfs
discourages the notion that photochemically induced aerosols are the
absorbing agent.  Absorption by warm dust or pressure-broadened
\ion{K}{1} remain viable and observationally testable hypotheses,
however.

Given the similarity of the colors and spectra of SDSS~1346--00,
SDSS~1624+00, and Gliese~229B, it is reasonable to assume that these
three brown dwarfs have similar luminosity.  Using this argument and
the measured distance to Gliese~229B of 5.8~pc (\cite{Hipparcos}),
Strauss et al.\ (1999) estimated a distance to SDSS~1624+00 of 10
pc. The apparent magnitude differences between SDSS~1346--00 and
SDSS~1624+00 are $\Delta{m}$ = +0.26 ($z'$), +0.29 ($J$), +0.28 ($H$),
and +0.14 ($K$).  The average difference (excluding $K$) of +0.3~mag
puts SDSS~1346--00 at a distance of 11.5~pc.  This estimate must be
treated with caution, however, since the SDSS~1346--00's larger flux
around 2.1~\micron\ may reflect a slightly higher temperature (and
hence luminosity as the models indicate that the radii of these
objects are essentially independent of the temperature or mass) than
those of Gliese~229B and SDSS~1624+00.

The SDSS commissioning data obtained to date cover approximately
400~deg$^2$, or $\sim~1$\%, of the sky.  To boost our confidence in
the one-band detections at faint magnitudes, we have searched only the
twice-imaged area of the sky for objects with $z^* \le 19.8$ ($\sim$
12$\sigma$ detection) and $i^*-z^* > 2.5$.  This strategy restricts
the searched area of the survey to 130~deg$^2$.  The two reddest
candidates in this restricted area, SDSS~1346--00 and SDSS~1624+00,
have been spectroscopically identified as methane brown dwarfs.
Recognizing the danger of statistical inferences based on a sample of
two objects, we estimate 635 such objects on the sky (of which
$\sim$~1/4 will be discovered by SDSS because of its sky coverage)
that satisfy our photometric-search criteria.  This implies a surface
density of 0.015~deg$^2$.  Using our detection limit of $z^* = 19.8$,
our search area of 130~deg$^2$, and Gliese~229B as a standard candle,
we estimate our search volume to be $\sim 40$~pc$^3$ and the space
density of Gliese~229B-like brown dwarfs to be 0.05~pc$^{-3}$.

Our surface-density estimate is $\sim 3$ times larger than that
derived by Strauss et al.\ (1999).  This discrepancy is due to our
reduction by 68\% of the search area and the doubling of the number of
detected methane dwarfs.  Based on the four objects identified from
1784~deg$^2$ of 2MASS, Burgasser et al.\ (1999) estimate $\sim 90$
``T'' dwarfs on the sky brighter than a $10~\sigma$ detection limit of
$J = 16$.  This number corresponds to a surface density of
0.0022~deg$^{-2}$ and a space density of $\sim 0.01$~pc$^{-3}$.  For
brown dwarfs with colors like those of SDSS~1346--00, the 2MASS
detection limit is equivalent to $z^* \approx 19.5$, {\it i.e.},
slightly brighter than our selection criterion of $z^* < 19.8$.
Despite the nearly equal sensitivity of 2MASS and SDSS to such brown
dwarfs, our estimates of surface and space density are larger than the
2MASS estimates by factors of $\sim 7$ and $\sim 5$, respectively.
Cuby et al.\ (1999) infer a space density of 1~pc$^{-3}$ from one
confirmed methane brown dwarf in the 2\farcm3~$\times$~2\farcm3 NTT
Deep Field.  This value is $\sim 20$ times higher than our estimate
for the same type of object.  The large dispersion in the SDSS, 2MASS,
and NTT estimates is almost certainly a result of small-sample
statistics.  We look forward to the imminent routine operation of SDSS
as a means of improving these very preliminary statistics.

The Sloan Digital Sky Survey (SDSS) is a joint project of the
University of Chicago, Fermilab, the Institute for Advanced Study, the
Japan Participation Group, The Johns Hopkins University, the
Max-Planck-Institute for Astronomy, Princeton University, the United
States Naval Observatory, and the University of Washington.  Apache
Point Observatory, site of the SDSS, is operated by the Astrophysical
Research Consortium.  Funding for the project has been provided by the
Alfred P. Sloan Foundation, the SDSS member institutions, the National
Aeronautics and Space Administration, the National Science Foundation,
the U.S. Department of Energy, and the Ministry of Education of Japan.
The SDSS Web site is {\tt http://www.sdss.org/}. We also thank Karen
Gloria for her expert assistance at the Apache Point Observatory.
UKIRT is operated by the Joint Astronomy Centre on behalf of the
U.K. Particle Physics and Astronomy Research Council. We are grateful
to the staff of UKIRT for its support, to A.\ J.\ Adamson for use
of UKIRT Director's time and to Tom Kerr for obtaining the medium 
resolution $J$-band spectrum.

\begin{center}
Table 1. Photometry of SDSS~J134646.45--003150.4
\begin{scriptsize}
\begin{tabular}{cccccccc}\\ \hline \hline
$u^*$ & $g^*$ & $r^*$ & $i^*$ & $z^*$ & $J$ & $H$ & $K$ \\ \hline
$24.11 \pm 0.38$ & $24.20 \pm 0.40$ & $24.54 \pm 0.62$ & $23.26 \pm 0.65$ & 
$19.29 \pm 0.06$ & $15.82 \pm 0.05$ & $15.85 \pm 0.05$ & $15.84 \pm 0.07$ \\
$24.08 \pm 0.39$ & $24.27 \pm 0.43$ & $24.07 \pm 0.42$ & 
$23.58 \pm 0.75$ & $19.23 \pm 0.08$ & & & \\ \hline 
\end{tabular}
\end{scriptsize}
\end{center}

\begin{figure}
\vspace{-0.5cm}
\epsfysize=600pt \epsfbox{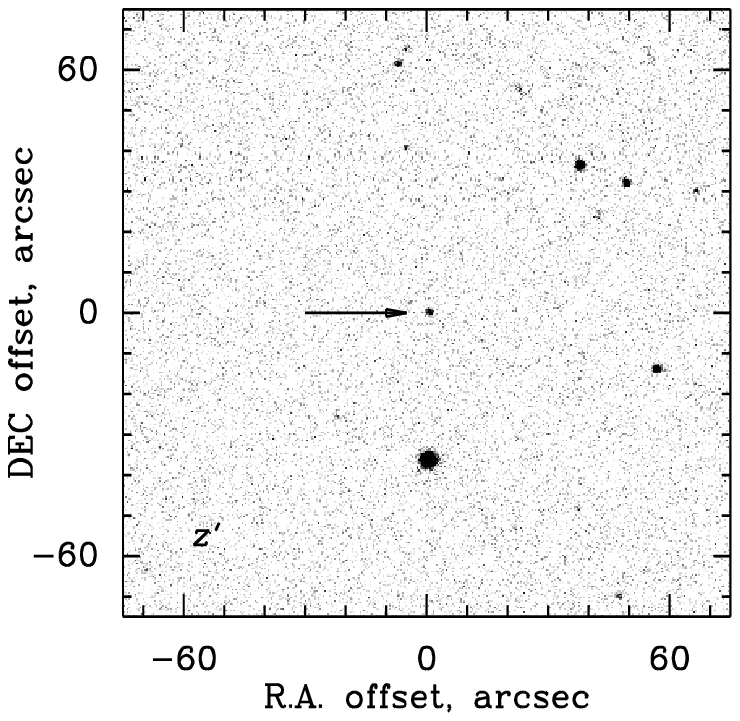}
\vspace{-1.0cm} Figure 1. Finding chart for SDSS~1346--00.  The panel
shows a section of the $z'$-band SDSS scan obtained on UT 1999
March~22.
\label{fig:finding-chart}
\end{figure}
\newpage

\begin{figure}
\vspace{-0.5cm}
\epsfysize=500pt \epsfbox{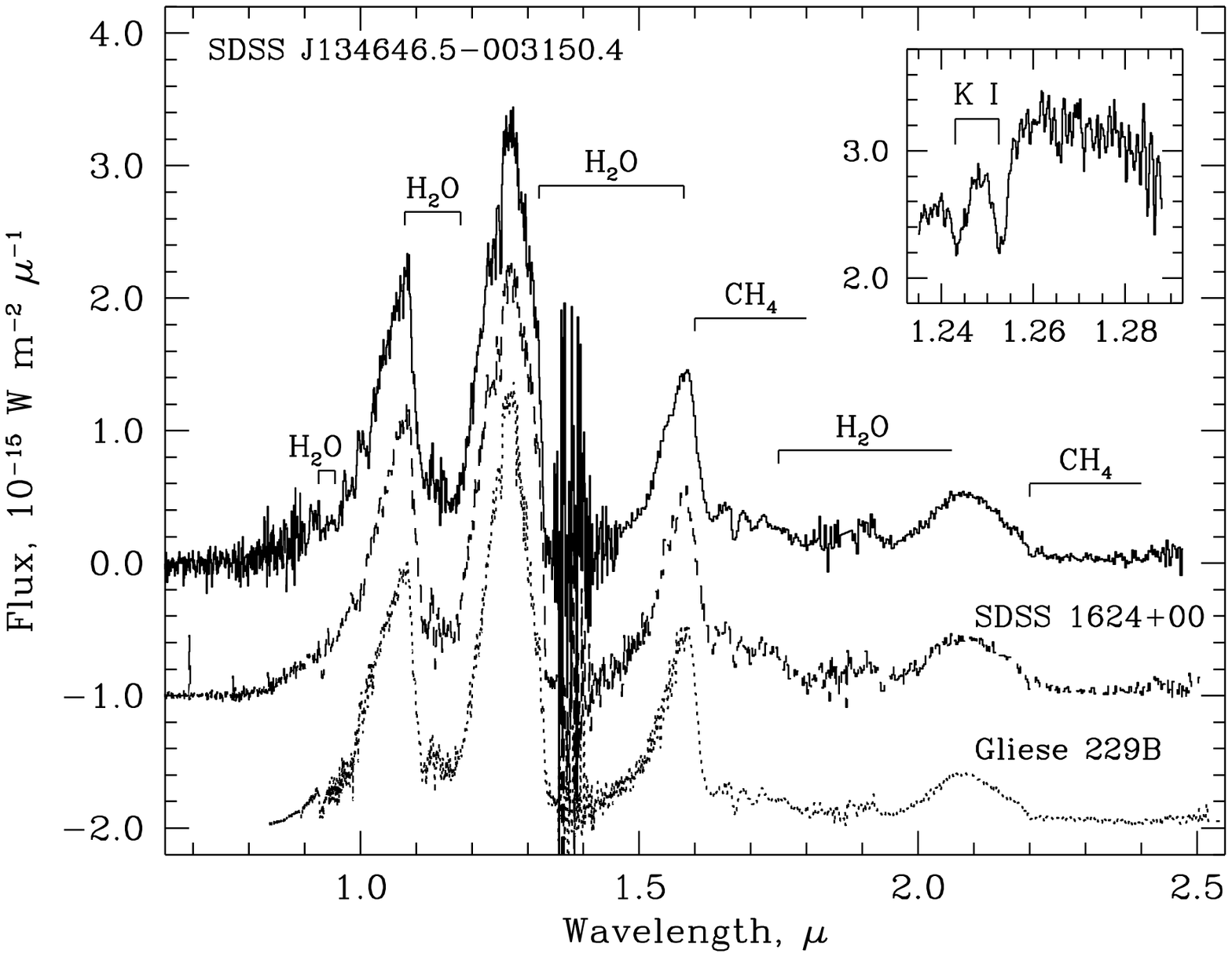}
\vspace{-2.0cm} Figure 2. The spectrum of SDSS~1346--00 from 0.8 to
2.5~\micron.  The prominent molecular bands of H$_2$O and CH$_4$ are
marked.  The spectra of SDSS~1624+00 and Gliese~229B are scaled to
match the peak of SDSS~1346--00 at $\sim$~1.27~\micron\ and then
offset for ease of comparison.  The scale factors and offsets are
(0.79,$-$1) and (0.24,$-$2), respectively. The {\it inset} shows a
medium resolution ($R \sim 3000$) spectrum centered about the peak of
the $J$-band emission.  Absorption lines from the \ion{K}{1} doublet
at 1.2436~$\mu$m and 1.2536~$\mu$m are well resolved.
\label{fig:sdss1346-spectrum}
\end{figure}
\newpage

\begin{figure}
\vspace{-0.5cm}
\epsfysize=500pt \epsfbox{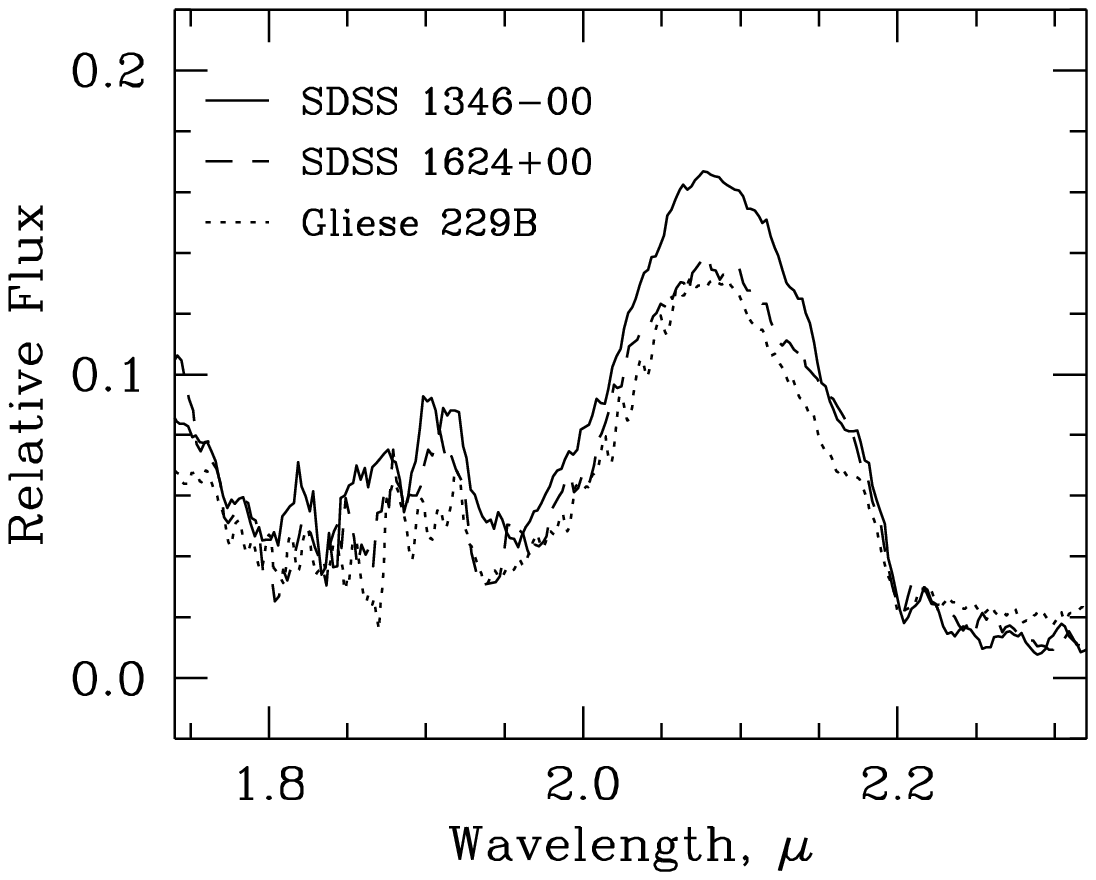}
\vspace{-2.0cm} Figure 3. Comparison of $K$-band spectra of
SDSS~1346--00 (solid), SDSS~1624+00 (dashed), and Gliese 229B
(dotted).  The three spectra are normalized to the peak of the
emerging flux around $\sim 1.27$~\micron\ and smoothed with a boxcar
of 5 pixels.
\label{fig:K-band-spectra}
\end{figure}

\end{document}